\documentclass[apsrev4-1,prl,amsmath,amssymb,superscriptaddress,twocolumn,longbibliography]{revtex4-1}
\usepackage{colortbl}
\usepackage{color}
\usepackage{amsfonts,amsmath,amssymb}
\usepackage[dvipsnames]{xcolor}
\usepackage{hyperref,tikz}
\usepackage{tabularx,hhline}
\usepackage{bm}
\usepackage{dcolumn,stmaryrd}

\begin{document}

\title{Spin Ice Thin Film:\\ Surface Ordering, Partial Magnetic Wetting and Emergent Square Ice}

\author{L. D. C. Jaubert}  
\affiliation{Okinawa Institute of Science and Technology Graduate University, Onna-son, Okinawa 904-0495, Japan}
\author{T. Lin}
\affiliation{Department of Physics and Astronomy, University of Waterloo, 200 University Avenue West, Waterloo, Ontario, N2L 3G1, Canada}
\author{T. S. Opel}
\affiliation{Department of Physics and Astronomy, University of Waterloo, 200 University Avenue West, Waterloo, Ontario, N2L 3G1, Canada}
\author{P. C. W. Holdsworth}
\affiliation{Universit\'e de Lyon, Laboratoire de Physique, \'Ecole normale sup\'erieure de Lyon, CNRS, UMR5672, 46 All\'ee d'Italie, 69364 Lyon, France}
\author{M. J. P. Gingras}
\affiliation{Department of Physics and Astronomy, University of Waterloo, 200 University Avenue West, Waterloo, Ontario, N2L 3G1, Canada}
\affiliation{Perimeter Institute for Theoretical Physics, 31 Caroline North, Waterloo, Ontario, N2L 2Y5, Canada}
\affiliation{Canadian Institute for Advanced Research, Toronto, Ontario, M5G 1Z8, Canada}
\date{\today}

\begin{abstract}

Motivated by recent realizations of Dy$_{2}$Ti$_{2}$O$_{7}$ and Ho$_{2}$Ti$_{2}$O$_{7}$ spin ice thin films, and more generally by the physics of confined gauge fields, we study a model of spin ice thin film with surfaces perpendicular to the $[001]$ cubic axis. The resulting open boundaries make half of the bonds on the interfaces inequivalent. By tuning the strength of these inequivalent ``orphan'' bonds, dipolar interactions induce a surface ordering equivalent to a two-dimensional crystallization of magnetic surface charges. This surface ordering can also be expected on the surfaces of bulk crystals. In analogy with partial wetting in soft matter, spins just below the surface are more correlated than in the bulk, but \emph{not} ordered. For ultrathin films made of one cubic unit cell, once the surfaces are ordered, a square ice phase is stabilized over a finite temperature window, as confirmed by its entropy and the presence of pinch points in the structure factor. Ultimately, the square ice degeneracy is lifted at lower temperature and the system orders in analogy with the well-known $F$-transition of the $6$-vertex model.
\end{abstract}

\maketitle


Highly frustrated magnets have been shown to host an astonishing array of exotic many-body phenomena, taking us far from the conventional paradigms of collective magnetic behavior~\cite{Springer_book}. The formulation of the local frustrated constraints in terms of effective gauge fields has revolutionized our perspective on these systems in both the classical and quantum domains. Depending on the system, this gauge 
symmetry can take the form of electromagnetism~\cite{Isakov2004,Henley05a,Lawler2013,Hermele2004} with photon and magnetic-monopole excitations~\cite{Hermele2004,Banerjee08a,Benton2012,Castelnovo2008,Ryzhkin2005,Jaubert2009b,Morris2009,Fennell2009,Kadowaki09a}, or be similar to quantum chromodynamics~\cite{cepas11a} and linearized general relativity~\cite{benton16a}, as well as support phase transitions lying outside the Ginzburg-Landau-Wilson framework \cite{Jaubert08a,Powell2008,Charrier08a,Powell2011}.

As vividly exposed in classic texts on electromagnetism~\cite{jackson99a}, boundaries dramatically influence the behavior of gauge fields. It is therefore natural to ask what may be the role of boundary conditions in frustrated magnets described by gauge theories. Classical spin ice~\cite{Harris97a} presents itself as a nascent paragon to lay the foundations of such concepts. This is further motivated by the recent growth of spin ice thin films of  Dy$_{2}$Ti$_{2}$O$_{7}$~\cite{Bovo14a,Petrenko14a} and Ho$_{2}$Ti$_{2}$O$_{7}$~\cite{Leusink14a,Kukli14a}, along with the promising possibilities offered by spin ice heterostructures~\cite{Sasaki14a,She16a} and pyrochlore-iridates thin films~\cite{Hu12a,Yang14a,Bergholtz15a,Hu15b,Fujita15a,Fujita16a,Hwang16a}.

In this paper, we study the dipolar spin ice model in thin film (slab) geometry (slab-DSI) defined by Eq.~(\ref{eq:ham}) below. As illustrated in Fig.~\ref{fig:TF}, the slab geometry renders the nearest-neighbor bonds on the surface inequivalent: either a bond belongs to a ``bulk tetrahedron'', or it is an ``orphan bond'', belonging to a ``virtual tetrahedron'' which has been ``cleaved away'' at the interface. Each virtual tetrahedron may therefore carry a magnetic surface charge that can propagate to and from the surface into the bulk. The orphan bonds act as effective chemical potentials for surface charges, allowing the slab-DSI to go through a surface charge-ordering transition. This is monopole crystallization \cite{Brooks2014} in two-dimensions, driven by the magnetic Coulomb potential between surface charges. Most notably, the crystallization is limited to the microscopic surface layer with no penetration into the bulk. The thinnest slab where this phenomenology can be explored systematically is one cubic unit-cell thick, containing three layers of tetrahedra. Below the surface ordering temperature at $T_{\rm so}$, the central layer emerges in the form of a constrained square ice system, which supports a state with extensive degeneracy described by a two-dimensional Coulomb phase~\cite{Henley2010}. Dipolar interactions ultimately break the symmetry of the Coulomb phase in the same way as in the venerable antiferroelectric $F$-model~\cite{Lieb67b}.\\

\begin{figure}
\centering\includegraphics[width=7cm]{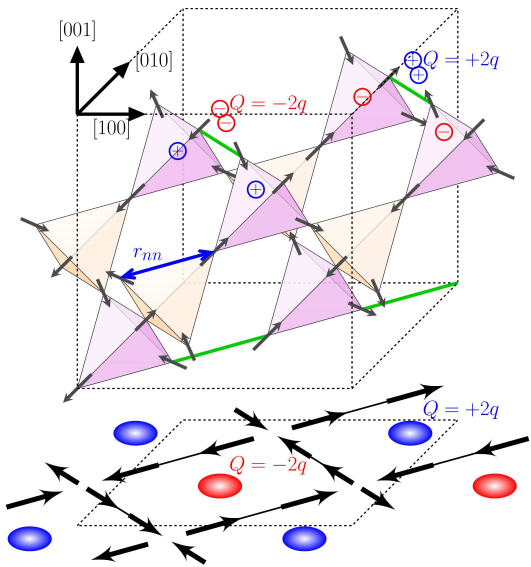}
\caption{Spin ice thin film of thickness $L_{h}=1$ unit cell in the $[001]$ cubic direction. Open boundaries lead to orphan bonds (green) on the surfaces. Each spin can be seen as a pair of $\textcolor{blue}{\oplus}$ and $\textcolor{red}{\ominus}$ effective  magnetic charges $(\pm q)$~\cite{Castelnovo2008}. Negative $J_o$ favors magnetic surface charges $Q=\pm 2 q$ on orphan bonds. Once the surface charges order in a checkerboard pattern for $T<T_{\rm so}$, the middle layer of tetrahedra (orange)  develop into a square ice phase, as seen by projection of the spins in plane. Here we show one of the ground states below the $F$-transition ($T<T_{F}$). $r_{nn}$ is the nearest-neighbor distance.}
\label{fig:TF}
\end{figure}

\textbf{Dipolar spin ice thin film --} We consider thin films of the pyrochlore lattice, made of corner-sharing tetrahedra, whose surfaces are normal to the [001] cubic axis [Fig.~\ref{fig:TF}]. We take the magnetic moments to be described by classical Ising pseudospins ${\bm S}_{i} \equiv\sigma_{i} \;\hat z_{i}$~\cite{Rau_quantum}, where $\hat z_{i}$ is the local easy-axis and $\sigma_{i}=\pm 1$~\footnote{The single-ion crystal field doublets of Dy$^{3+}$ and Ho$^{3+}$ are defined through a spectral weight in which $m_J=\pm J$ dominate and whose angular momenta are $J=15/2$ and $J=8$, respectively. As $m_J$ is maximal, we may assume that the Ising nature of the moment at the surface remains largely protected through the absence of high rank crystal field perturbations that would renormalize the doublet splitting~\cite{Rau_quantum}.}. The spins interact via nearest-neighbor couplings $J$ and $J_{o}$ and long-range dipolar interactions $D$:
\begin{eqnarray}
\mathcal{H}&=& J\sum_{\langle ij\rangle}\; \sigma_{i}\; \sigma_{j} \;+\; J_{o}\sum_{\langle ij\rangle_{\rm orphan}}\; \sigma_{i}\; \sigma_{j} \nonumber\\
&+& D\, r_{nn}^{3}\;\sum_{i>j} \frac{\hat z_{i}\cdot \hat z_{j} \, -\, 3 \left(\hat z_{i}\cdot\hat e_{ij}\right)\left(\hat z_{j}\cdot \hat e_{ij}\right)}{r_{ij}^{3}}\sigma_{i}\; \sigma_{j},
\label{eq:ham}
\end{eqnarray}
where the second sum runs over surface orphan bonds only [Fig.~\ref{fig:TF}].
$r_{nn}$ is the nearest-neighbor distance and $\hat e_{ij}$ is a unit vector between sites $i$ and $j$.

The bulk DSI model is characterized by an extensive low energy band of states \cite{Gingras01a,Isakov05a} in which each tetrahedron satisfies the ice rules with two spins pointing inwards and two spins pointing outwards (2-in/2-out)~\cite{Harris97a}. 
This is a Coulomb phase with Pauling entropy~\cite{Henley2010}. Topological excitations out of the Coulomb phase carry either a single (3-in/1-out or 3-out/1-in) or a double (4-in or 4-out) gauge charge. Dipolar interactions dress these topological excitations with an effective magnetic charge $Q$. This is the dumbbell model where each moment is recast as a pair of positive/negative magnetic charges $\pm q$ sitting at the centers of the adjoining tetrahedra [Fig.~\ref{fig:TF}]~\cite{Castelnovo2008}. Within the dumbbell model, the low energy band is exactly degenerate. The ordering at low-temperature in bulk-DSI~\cite{Melko01a,Melko04a} is due to corrections to this dumbbell description~\cite{Isakov05a,Castelnovo2008}.

Slab-DSI differs from bulk-DSI~\cite{gingras11a} because of the open boundaries in the [001] direction and the presence of orphan bonds. Nevertheless, the dumbbell model remains a useful description if one also considers surface charges [Fig.~\ref{fig:TF}]. To set an experimental context we use $J=-1.24$ K and $D=1.41$ K, as in a minimal model of Dy$_{2}$Ti$_{2}$O$_{7}$~\cite{Hertog00a}.


\textbf{Method --} Our approach is based on Monte Carlo simulations with parallel tempering~\cite{swendsen86a,geyer91a} and loop algorithm. The later has been adapted to include dipolar interactions~\cite{Melko01a,Melko04a} and the presence of monopoles~\cite{Brooks2014,Lin14a}. The system size is $L\times L\times L_h$ in units of the 16-site cubic unit cell, with thickness $L_{h}<L$. We take open boundary conditions along the [001] cubic axis and periodic ones along the other two cubic axes. The dipolar energy is computed by the Ewald summation in slab geometry without a demagnetization factor~\cite{Yeh99a,Brodka04a,Melko04a}. In previous numerical studies of spin ice, the Ewald summation was implemented using periodic replication of the simulated system in the three cubic directions. However, it is straighforward to reproduce the slab geometry by inserting an empty space, as large as needed, between replicas in the [001] direction~\cite{Yeh99a,Brodka04a}. We chose an empty space of $L_{e}=1000$ unit cells. No difference was observed over the temperature range considered upon varying $L_{e}$ for $L_{e}\gg L>L_{h}$.
\\

\begin{figure}
\centering\includegraphics[width=\columnwidth]{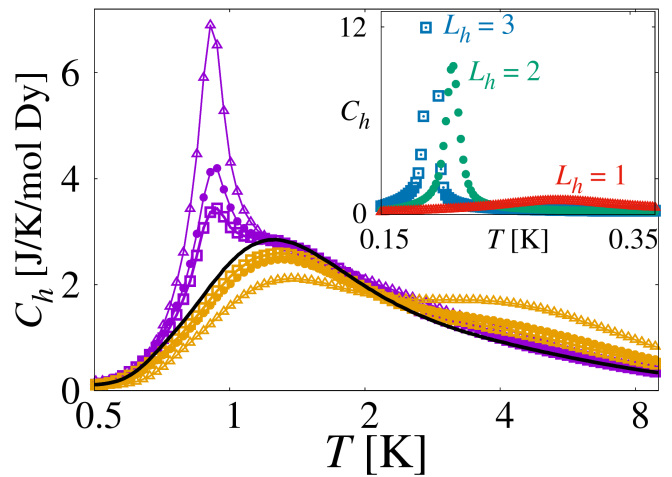}
\caption{
Specific heat $C_h $ for thin films ($J_{o}=+ 4$ K in orange and $-4$ K in violet) of different thickness ($L_h=1$ ($\blacktriangle$), 2 ($\bullet$), 3 ($\boxdot$)) and for bulk dipolar spin ice (bulk-DSI, black). \textit{Inset:} Low temperature $C_h $ for $J_o=-4$ K. The temperature axes are in log (main) and linear (inset) scales. The system sizes are $L=8$ (main) and $L=6$ (inset). Throughout the paper $J=-1.24$ K and $D=1.41$ K. With these $J$ and $D$ values, the low-temperature $T_F$ transition approaches 
the bulk $T_c\approx 0.18$ K~\cite{Melko01a,Melko04a} in the limit $L_h \rightarrow \infty$.}
\label{fig:Ch}
\end{figure}
\begin{figure}
\centering\includegraphics[width=0.9\columnwidth]{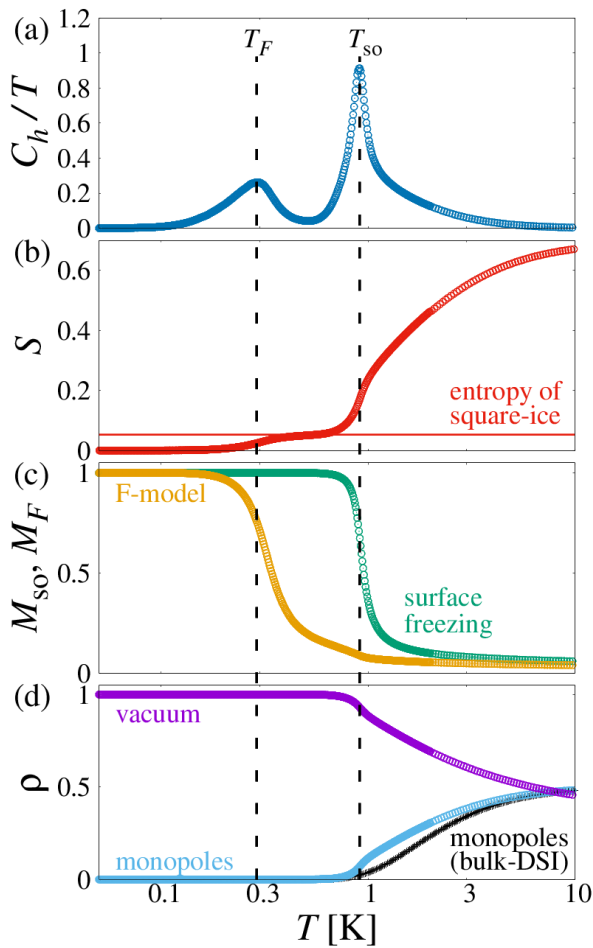}
\caption{Two-step ordering in thin films of thickness $L_{h}=1$ and $J_{o}=-4$ K: (a) specific heat, (b) entropy, (c) order parameters and (d) densities of 2-in/2-out tetrahedra (violet) and single charges (blue) inside the thin film, \textit{i.e.} not on the surfaces. The latter is noticeably higher than the density of single charges in the bulk-DSI model (black). The surface ordering at $T_{\rm so}\approx 900$ mK stabilizes a square ice model in the middle of the slab, as confirmed by the entropy in the intermediate region (delimited by the dashed lines). Dipolar interactions ultimately lift the square-ice entropy at $T_{F}\approx 300$ mK (orange) in favor of the same antiferromagnetic ground state as the $F$-model~\cite{Lieb67b}. The temperature, $T$, is in log scale.}
\label{fig:obs}
\end{figure}

\textbf{Surface ordering --} 
By varying $J_{o}$ with respect to a threshold value $J_{o}^c=-J-1.945(2) D$, one can favor either antiferromagnetic or ferromagnetic  configurations on orphan bonds. For $J_{o}>J_{o}^c$, this extra surface energy scale manifests itself in the specific heat as a  Schottky peak at $T \sim J_{o}$, with $J_{o}=+4$ K in Fig.~\ref{fig:Ch}. Bulk physics is then rapidly recovered as $L_{h}$ increases. For $J_{o} < J_{o}^c$, a sharp feature appears in the specific heat at an intermediate temperature $T_{\rm so}$ whose height decreases with increasing $L_{h}$ [$J_{o}=-4$ K in Fig.~\ref{fig:Ch}], suggesting a surface rather than a bulk transition, as discussed below.

One can think of the difference between $J$ and $J_{o}$ as a chemical potential shift for surfaces charges, compared to the bulk. Indeed, the ferromagnetic (resp.~antiferro.) alignment of the pseudospins on an orphan bond amounts to a net magnetic charge (resp.~vacuum) placed above the orphan bond at its mid point [Fig.~\ref{fig:TF}]. The charge sites form a square array with a lattice constant of $2 r_{nn}$. The dipolar interaction between spins generates an effective Coulomb interaction between the surface charges~\cite{Castelnovo2008}, so that each surface of the thin film can be thought of as a (square) Coulomb lattice gas. The surface charges can propagate into the bulk, which increases the density of monopoles above that of the bulk-DSI at intermediate temperature  [Fig.~\ref{fig:obs}.(d)]. At low temperature, when varying $J_{o}$ below $J_{o}^c$, the surface state transforms from charge vacuum to charge crystal with a checkerboard pattern in order to minimize the Coulomb potential [Fig.~\ref{fig:TF} and Fig.~\ref{fig:obs}(c)]. A suitable order parameter is $M_{\rm so}=|M^{b}|+|M^{t}|$ and $M^{b,t}=\sum_{\alpha} Q^{b,t}_{\alpha}\,\epsilon_{\alpha}/(2L^{2})$, where $Q^{b,t}_{\alpha}$ is a bottom ($b$) or top ($t$) surface charge and $\epsilon_{\alpha}=\pm 1$ accounts for the bipartite nature of the square lattice. Most importantly such surface order is not limited to thin films and could also occur on a sufficiently pristine $[100]$ surface of bulk crystals. The enhancement of monopole density at intermediate temperature should also persist 
within a thin layer below the surface of spin ice compounds,
which could then be manipulated by a magnetic field~\cite{Jaubert2009b}.
Near-surface dynamical properties could possibly be probed by $\beta$-NMR~\cite{beta-NMR1,beta-NMR2} or $\mu$SR with slow muons~\cite{slow_muons}.
\\

\begin{figure}
\includegraphics[width=\columnwidth]{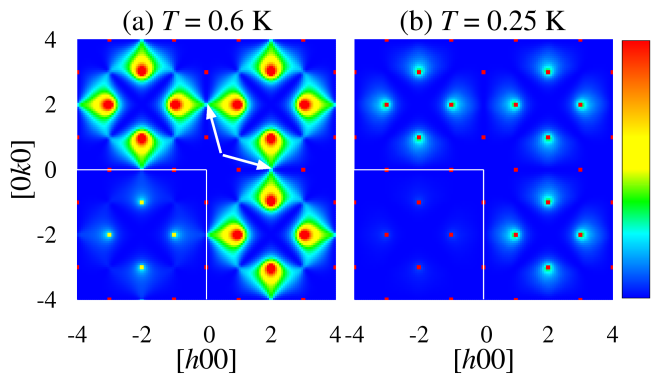}
\caption{(a,b) Structure factor $\mathcal{S}(\bm q)$ in the [hk0] Fourier plane for $L_{h}=1$ and $J_{o}=-4$ K. Square-ice physics gives rise to pinch points in the [200] (and symmetry equivalent) Brillouin zone centers~\cite{youngblood81a} [see white arrows]. These pinch points are (a) clear singularities at $T=600$ mK, but (b) disappear below the $F$-transition. The insets are the same data plotted on a broader color scale, in order to emphasize that the $\mathbf{q}=$[$\bar{2}\bar{1}0$] features are (b) Bragg peaks of the low-temperature long-range order, but (a) \textit{not} in the intermediate regime, thus confirming the two-step ordering. The color scale is the same for all three main panels and insets respectively.
}
\label{fig:SQ}
\end{figure}

\textbf{Emergent square ice for $\bm{L_{h}=1}$ --} Below $T_{\rm so}$, the system enters a temperature regime in which a large majority of the tetrahedra respect the ice rules. Considering a tetrahedron on the top layer, the orientation of the two surface spins becomes fixed by the surface ordering [Fig.~\ref{fig:TF}]. The two lower spins are not fixed, but are coupled by the ice rules, forming a composite $\mathbb{Z}_{2}$ degree of freedom with projection along $\pm$[110]. The same holds for the tetrahedra on the bottom layer, with composite spins projected along $\pm$[$\bar{1}$10]. These composite projections now form the famous $6$-vertex model~\cite{baxter07a} whose vertices correspond to the middle-layer tetrahedra respecting the ice rules [Fig.~\ref{fig:TF}]. This is confirmed by an entropy plateau for intermediate temperature at $S_{\rm sq}=\frac{1}{4}\,\left(\frac{3}{4}\,\ln\frac{4}{3}\right)$ (Fig.~\ref{fig:obs}(b)), which is the exact square ice entropy~\cite{lieb67c} (the composite spins account for $1/4$ of the original degrees of freedom). The vertical ($[001]$) heterogeneity of this phase is manifest in the pseudospin structure factor, $\mathcal{S}(\bm q) \equiv \frac{1}{N}\left|\sum_{j=1}^{N} \sigma_{j} e^{i \bm q \cdot \bm r_{j}}\right|^{2}$. $\mathcal{S}(\bm q)$ shows the co-existence of pinch points characteristic of the square ice phase, and Bragg peaks due to the surface charge order [Fig.~\ref{fig:SQ}(a)].\\

\textbf{Ground state -- }
Just as in bulk-DSI~\cite{Melko01a,Melko04a}, the degeneracy of the Coulomb phase is eventually lifted at sufficiently low temperature, here $T_F$, to give a long-range ordered ground state [Figs. \ref{fig:TF} and \ref{fig:Ch} (inset)]. However, the 12-fold degenerate ground states of bulk-DSI are inequivalent in slab geometry, 
as only 4 of them support charge order on the (001) surfaces. For these 4 states, if one flips all spins on either one or both surfaces, the new configurations are quasi-degenerate ground states when $J_o<J_o^c$. The breadth of this quasi-degeneracy is extremely small ($<0.05\%$), and decreases rapidly for increasing $L_{h}$. When $J_o>J_o^c$, orphan bonds favor 4 of the bulk-DSI ground states with zero surface charges.

When $L_{h}=1$, the transition at $T_F$ involves the square-ice degrees of freedom and is equivalent to the $\mathbb{Z}_{2}$ symmetry breaking of the antiferroelectric $F$-model~\cite{Lieb67b} [Figs.~\ref{fig:TF} and \ref{fig:obs}(c)]. The order parameter is $M_{F}=\sum_{i} \sigma_{i}\,\eta_{i}/N'$ where $\eta_{i}=\pm 1$ transcribes the $\mathbb{Z}_{2}$ ordering of the $N'=8 L^{2}$ spins of the middle layer of tetrahedra [Fig.~\ref{fig:TF}]. As a consequence, the ordering of the ``bulk'' of the thin film falls in a different universality class for $L_h=1$ and for $L_h>1$, as illustrated by the evolution of the specific heat in the inset of Fig.~\ref{fig:Ch}. The transition is characterized
in $\mathcal{S}(\vec q)$, by the development of new Bragg peaks and the extinction of the pinch points associated with the square ice Coulomb phase at $T_F$ [Fig.~\ref{fig:SQ}(b)].\\

\textbf{Discussion -- } In this work, we provided a benchmark for spin ice thin films. We showed that this model system is characterized by a partial magnetic wetting phenomenon: the lack of penetration of the surface phase into the bulk. It occurs here as a direct consequence of the spin liquid nature of bulk spin ice, so that bulk physics is rapidly recovered as one increases the film thickness [Fig.~\ref{fig:Ch}]. This is true, independently of the surface conditions in the form of orphan bonds, while the observed surface phase transition is a rare example of surface ordering rather than surface melting. \footnote{Surface ordering, does occur in other situations, for example, in free standing liquid crystal films, in alkaline chain polymers of intermediate length and gallium alloys.}. We are therefore exposing here a microscopic mechanism that generates frozen boundary conditions and hence could possibly implement a $3$-dimensional variant of the theoretical concept of ``arctic circle''~\cite{arctic_circle}. What we observe is at odds with recent experiments on Dy$_{2}$Ti$_{2}$O$_{7}$ thin films where no residual Pauling entropy was observed for a broad range of film thickness~\cite{Bovo14a}. Our results suggest that the explanation for these intriguing experiments may lie in in the presence of long range strain fields, as proposed in Ref.~[\onlinecite{Bovo14a}], and/or confinement induced by random disorder,  a topic of growing interest in pyrochlores~\cite{Ross12a,Taniguchi13a,Sala14a}. Our results also highlight the importance of long-range interactions both in thin films and for the surfaces of clean bulk crystals.

It is interesting to compare in more detail the partial wetting observed here to that occurring in water ice. Ice crystals show a thin surface layer stabilized down to $-40^\circ$C with remarkably high charge mobility~\cite{Fletcher62a,Ryzhkin01a,Li07a,Ryzhkin09a}. Although often described as a liquid layer, its conductivity is orders of magnitude \textit{larger} than that of bulk liquid water. This is admittedly a vastly more complex problem than our model spin ice~\cite{Fletcher92a,Buch08a,Pan08a}.
However, on cleaving a surface off  an ice crystal, one might expect polarization charge to be induced in an analogous manner. While the putative charge crystal could then melt under the high Coulombic pressure,  it has also been proposed that the high surface conductivity could come through a super-ionic mechanism in an \emph{ordered} ionic array \cite{Ryzhkin01a}, reminiscent of the surface ordering we found here. In this context it could be interesting to investigate the surface charge (monopole) conductivity of our model in future studies.

Finally, we return to ultra-thin films and the emergence of an effective square ice system over a finite temperature range for $L_h=1$. The obvious candidates 
for such  behavior are artificial spin ice (ASI) arrays~\cite{Wang06a}. However, they  fail miserably on this score \cite{Moller06a} as, with planar geometry, the band width of ice rules states is as large as the energy scale for producing topological defects. The band width can be reduced in model systems by varying the relative height between horizontal and vertical chains of nano-islands~\cite{Moller06a,Chern14c}. This is naturally what happens here with the pyrochlore geometry~\cite{Hertog00a,Isakov05a} which, in addition, by having magnetic moment orientations along the $[111]$ axes of the cubic cell, possesses the high symmetry of the vertex configurations that ASI arrays lack. This higher symmetry pushes the $F$-transition to lower temperature,  [Fig.~\ref{fig:obs}], offering a  solid state realization for such  two-dimensional spin ice physics.

In conclusion, we believe the theory presented here is just the tip of the iceberg of possibilities for spin ice films and surfaces of single crystals. We hope our work will motivate further efforts in the investigation of surface and confinement phenomena in frustrated magnetism and other strongly correlated systems described by an emergent gauge theory.\\

\begin{acknowledgments}
We thank Bruce Gaulin, Gabriele Sala, Etienne Lantagne-Hurtubise, Rob Kiefl, Andrew MacFarlane, Jeff Rau, Nic Shannon and Tommaso Roscilde for useful discussions. We acknowledge the hospitality of the Ecole Normale Sup\'erieure de Lyon, Universit\'e Lyon 1 and CNRS (LDCJ and MJPG) and from the Okinawa Institute of Science and Technology Graduate University (PCWH and MJPG). LDCJ is supported by the Okinawa Institute of Science and Technology Graduate University. The work at the U. of Waterloo was supported by the NSERC of Canada, the Canada Research Chair program (M.G., Tier 1) and by the Perimeter Institute (PI) for Theoretical Physics. Research at the Perimeter Institute is supported by the Government of Canada through Innovation, Science and Economic Development Canada and by the Province of Ontario through the Ministry of Research, Innovation and Science.
\end{acknowledgments}
\bibliographystyle{apsrev4-1}
\bibliography{Thin-film}
\end{document}